\PassOptionsToPackage{unicode=true}{hyperref} 
\PassOptionsToPackage{hyphens}{url}
\documentclass[]{article}
\usepackage{multirow}
\usepackage{lmodern}
\usepackage{amssymb,amsmath}
\usepackage{ifxetex,ifluatex}
\usepackage{fixltx2e} 
\ifnum 0\ifxetex 1\fi\ifluatex 1\fi=0 
  \usepackage[T1]{fontenc}
  \usepackage[utf8]{inputenc}
  \usepackage{textcomp} 
\else 
  \usepackage{unicode-math}
  \defaultfontfeatures{Ligatures=TeX,Scale=MatchLowercase}
\fi
\IfFileExists{upquote.sty}{\usepackage{upquote}}{}
\IfFileExists{microtype.sty}{%
\usepackage[]{microtype}
\UseMicrotypeSet[protrusion]{basicmath} 
}{}
\IfFileExists{parskip.sty}{%
\usepackage{parskip}
}{
\setlength{\parindent}{0pt}
\setlength{\parskip}{6pt plus 2pt minus 1pt}
}
\usepackage{hyperref}
\hypersetup{
            pdfborder={0 0 0},
            breaklinks=true}
\usepackage{longtable,booktabs}
\IfFileExists{footnote.sty}{\usepackage{footnote}\makesavenoteenv{longtable}}{}
\usepackage{graphicx,grffile}
\makeatletter
\def\maxwidth{\ifdim\Gin@nat@width>\linewidth\linewidth\else\Gin@nat@width\fi}
\def\maxheight{\ifdim\Gin@nat@height>\textheight\textheight\else\Gin@nat@height\fi}
\makeatother
\setkeys{Gin}{width=\maxwidth,height=\maxheight,keepaspectratio}
\ifx\paragraph\undefined\else
\let\oldparagraph\paragraph
\renewcommand{\paragraph}[1]{\oldparagraph{#1}\mbox{}}
\fi
\ifx\subparagraph\undefined\else
\let\oldsubparagraph\subparagraph
\renewcommand{\subparagraph}[1]{\oldsubparagraph{#1}\mbox{}}
\fi

\makeatletter
\def\fps@figure{htbp}
\makeatother

\date{}

\usepackage{geometry}
\begin{document}

\hypertarget{tf3p-a-new-three-dimensional-force-fields-fingerprint-learned-by-deep-capsular-network}{%
\section{TF3P: A New Three-dimensional Force Fields Fingerprint Learned
by Deep Capsular
Network}\label{tf3p-a-new-three-dimensional-force-fields-fingerprint-learned-by-deep-capsular-network}}

Yanxing Wang\textsuperscript{†,1}, Jianxing Hu\textsuperscript{†,1},
Junyong Lai\textsuperscript{1}, Yibo Li\textsuperscript{2}, Hongwei
Jin\textsuperscript{1}, Lihe Zhang\textsuperscript{1}, Liang-Ren
Zhang\textsuperscript{*,1}, Zhen-ming Liu\textsuperscript{*,1}

\textsuperscript{1}: State Key Laboratory of Natural and Biomimetic
Drugs, School of Pharmaceutical Sciences, Peking University, Beijing,
100191, P. R. China

\textsuperscript{2}: Academy for Advanced Interdisciplinary Studies,
Peking University, Beijing, 100191, P. R. China

\textsuperscript{†}: These authors contributed equally.

\textsuperscript{*}: To whom correspondence should be addressed.

Author Email Address:

Yanxing Wang: yh\_wang@bjmu.edu.cn;

Jianxing Hu: j.hu@pku.edu.cn;

Junyong Lai: jylai@bjmu.edu.cn;

Yibo Li: ybli@pku.edu.cn;

Hongwei Jin: jinhw@bjmu.edu.cn;

Lihe Zhang: zdszlh@bjmu.edu.cn;

Liang-Ren Zhang: liangren@bjmu.edu.cn;

Zhen-ming Liu: zmliu@bjmu.edu.cn.

\hypertarget{abstract}{
\subsection{Abstract: }\label{abstract}}

Molecular fingerprints are the workhorse in ligand-based drug discovery.
In recent years, an increasing number of research papers reported
fascinating results on using deep neural networks to learn 2D molecular
representations as fingerprints. It is anticipated that the integration
of deep learning would also contribute to the prosperity of 3D
fingerprints. Here, we unprecedentedly introduce deep learning into 3D
small molecule fingerprints, presenting a new one we termed as the
{t}hree-dimensional {f}orce {f}ields {f}inger{p}rint (TF3P). TF3P is
learned by a deep capsular network whose training is in no need of
labeled datasets for specific predictive tasks. TF3P can encode the 3D
force fields information of molecules and demonstrates the stronger
ability to capture 3D structural changes, to recognize molecules alike
in 3D but not in 2D, and to identify similar targets inaccessible by
other 2D or 3D fingerprints based on only ligands similarity.
Furthermore, TF3P is compatible with both statistical models (e.g.
similarity ensemble approach) and machine learning models. Altogether,
we report TF3P as a new 3D small molecule fingerprint with a promising
future in ligand-based drug discovery. All codes are written in Python
and available at
\href{https://github.com/canisw/tf3p}{{https://github.com/canisw/tf3p}}.

\textbf{Keywords:} Deep Learning; Capsular Network; 3D Molecular
Fingerprint; Force Fields; Ligand-based Target Prediction.

\newpage{}
\hypertarget{introduction}{%
\subsection{Introduction}\label{introduction}}

It is a primary principle in ligand-based drug discovery that similar
ligands bind to similar targets\textsuperscript{1}. This statement, in
essence, is to infer target similarity from only ligand information.
Over the past decades of endeavor to improve this inference with more
precision, molecular representation is a useful
tool\textsuperscript{2-5} and can be divided intuitively into two
categories, namely 2D and 3D representations.

Since the 1970s, 2D molecular fingerprints have been developed maturely
and can be classified into four types\textsuperscript{3}, namely
substructure keys (e.g. MACCSKey\textsuperscript{6}), circular
fingerprints (e.g. ECFP\textsuperscript{7}), topological fingerprints
(e.g. atom pairs\textsuperscript{8}, topological
torsions\textsuperscript{9}, Avalon fingerprint\textsuperscript{10}) and
pharmacophores. In addition to these classical molecular fingerprints
designed by chemists, with the advent of deep learning in recent years,
an increasing number of research papers reported on using deep neural
networks (DNN) to fingerprint 2D molecules\textsuperscript{11-17}. Of
particular note is the recent Attentive FP\textsuperscript{15} which
achieves the state-of-art performance on several datasets in the field
of ligand-based bioactivity prediction.

Compared to 2D representation, using molecular 3D
representation\textsuperscript{18} is expected to enhance the
performance of predictive models, especially in the prediction of
biological targets for small molecule drugs.
USR/USRCAT\textsuperscript{19, 20} is moment-based molecular 3D
representations that encode molecular shape (and color for USRCAT).
E3FP\textsuperscript{21}, as another 3D molecular fingerprint, was
inspired by ECFP and presents higher precision-recall performance than
ECFP when integrated with the similarity ensemble approach
(SEA)\textsuperscript{22}. Besides the above rotation and translation
invariant fingerprints, Gaussian-based representation (used in
ROCS\textsuperscript{23} and SimG\textsuperscript{24}) is another
example that strengthens this argument. When it further comes to
voxelized representation\textsuperscript{25, 26}, it is a common
practice to integrate this representation with supervised convolutional
DNN and promising results have been demonstrated on the prediction of
the biological target or binding affinity for small
molecules\textsuperscript{27-32}. Despite all the above, there is no 3D
fingerprint learned by DNN reported thus far.

Whereas 2D molecular fingerprints generated by DNN showed excellent
ability in many predictive models, most of them reported up to now have
two flaws: i) Such fingerprinting models need individual training for
each predictive tasks, resulting in that none of generated fingerprints
contain the unbiased information within an intact molecule; ii)
Similarity calculation of such fingerprints is a complication to their
compatibility with similarity-based statistical models such as
SEA\textsuperscript{22}. In 2019, Winter et al. introduced a DNN-based
translation model between two equivalent chemical representations (e.g.
IUPAC name and SMILES) to deal with these two flaws\textsuperscript{33}.
While this model can be trained as a general fingerprinter with such
chemical representations and simultaneously nine more molecular
properties (i.e. log P, etc.), the fact that the generated
representation must be normalized over the dataset of interest before
similarity calculation defines its deficiency. Our research presented
here is also aimed to overcome these limitations and further push
forward with the neural network to fingerprint 3D molecules.

Fingerprinting 3D molecule is to encode 3D object. We can learn a lot
about the encoding method from computer graphics. Hinton et al. recently
introduced CapsNet\textsuperscript{34-36} to learn to output the pose
matrix of 3D objects, which can be intuitively interpreted as the
orientation and position of objects in 3D space. The idea behind CapsNet
is to surmount one deficiency of the convolutional neural network (CNN),
as shown in \textbf{Figure 1}. Each capsule in CapsNet is capable to
learn the instantiation parameters (i.e. values for orientation, size,
etc.) of each feature and output a vector with its length as the
possibility indicating the presence of that feature. Taken all features
together, CapsNet outputs a matrix that represents the input. It is
worth noting that capsules do not guarantee provable
rotation/translation equi-/in-variance intrinsically. In 2018 and late
2019, P. Libuschewski et al.\textsuperscript{37} and R. R. Sarma et
al.\textsuperscript{38} advanced the equivariance of capsular networks,
but there is still no remedy for invariance.

\includegraphics[width=5.99792in,height=1.18403in]{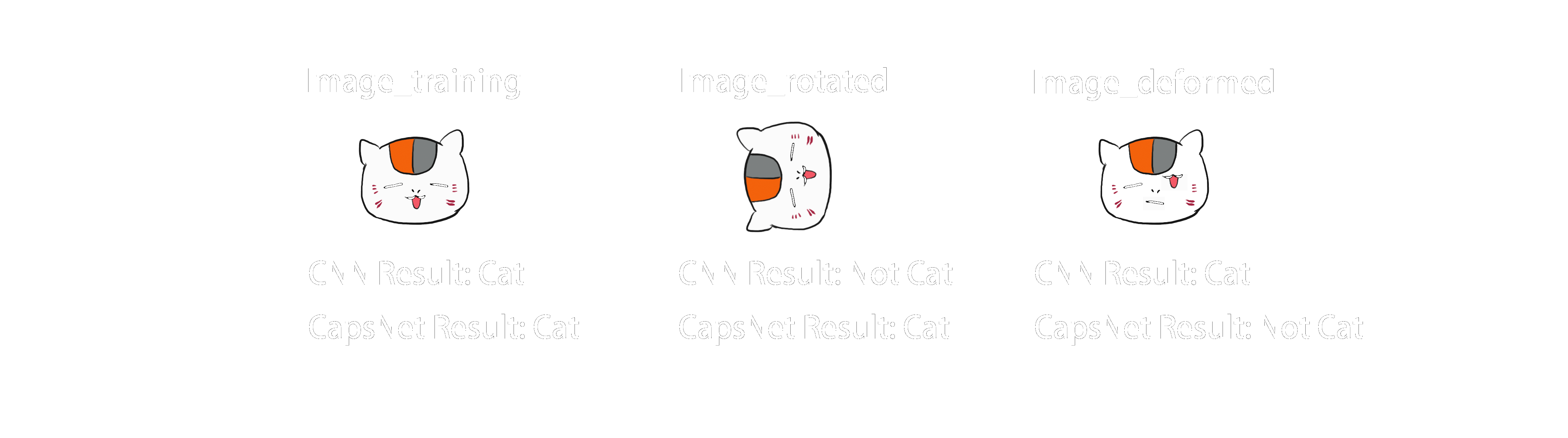}

\textbf{Figure 1}. The CNN cannot detect relationships between features
and then fails to detect deformed samples, but CapsNet can.

In this paper, we construct a deep capsular neural network to
fingerprint 3D small molecule, which can be trained without targeting
specific predictive tasks. Molecular force fields grids, as a kind of
voxelized representation employed in 3D field-based
QSAR\textsuperscript{39, 40}, were adopted as the inputs of our neural
network. This was in the very consideration that one intended
application of our fingerprint is ligand-based target prediction and
that the binding of small molecules to biological targets is dominated
by physical forces. All in all, molecular force fields grids can be
compressed with the well-trained model into a real-valued matrix, termed
as the {t}hree-dimensional {f}orce {f}ields {f}inger{p}rint (TF3P). We
further demonstrated TF3P's strong ability to capture 3D structural
changes and recognize similar targets which are inaccessible by other
fingerprints based on only ligands similarity. Additionally, TF3P is
compatible with both statistical models and machine learning models.

\hypertarget{methods-materials}{%
\subsection{Methods \& Materials}\label{methods-materials}}

\hypertarget{model-architecture}{%
\subsubsection{Model Architecture}\label{model-architecture}}

Our model is a modified version of CapsNet for 3D convolutions, composed
of 15 layers. The first 8 layers are encoder and the last 7 layers are
decoder. The overall architecture is shown in \textbf{Figure 2}.

\includegraphics[width=5.99722in,height=2.8375in]{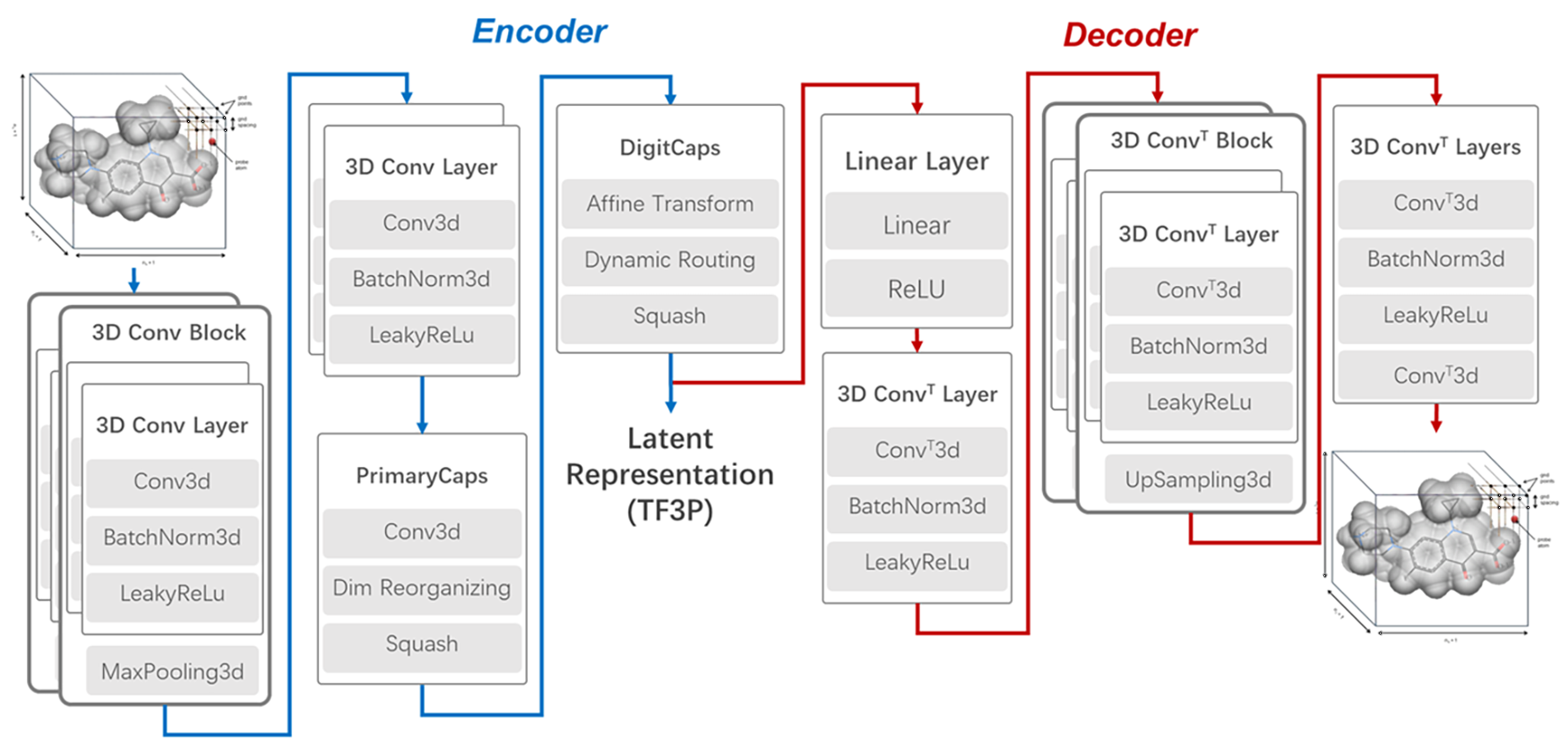}

\textbf{Figure 2.} The architecture of our model. The arrows indicate
the data flow. TF3P: the {t}hree-dimensional {f}orce {f}ields
{f}inger{p}rint.

\emph{Inputs.} Our model takes the force fields grids of a molecule's
conformer as inputs, which have two channels, namely van der Waals
potential and electrostatic potential. The grids calculation is
implemented using open source force field MMFF94\textsuperscript{41-45}
in Python with alkyl carbon (MMF type 1) and proton as probes without
solvents for each channel, respectively. The box size is set to
\(20 \times 20 \times 20\ \mathring{\mathrm{A}}\) so that it can hold
common lead-like small molecule drugs. The grids' size is set to
\(50 \times 50 \times 50\) points with a resolution of
\(0.4\ \mathring{\mathrm{A}}\). In summary, the input tensor for each
conformer has the shape of \(2 \times 50 \times 50 \times 50\).

\emph{Encoder.} This part of the network takes the input force fields
grids and learns to encode them into a \(166 \times 8\) matrix as the
fingerprint. The row vector dimension of 8 is chosen as the same value
used by Hinton et al. The encoder consists of six 3D convolutional
layers and two capsular layers. The convolutional layers serve as a
feature extractor to reduce the sparse inputs' dimension into a smaller
size suitable for the subsequent capsular layers since capsular layer
has hundred times of trainable parameters than convolutional layer under
the same scale. Every convolutional layer has kernels with a size of
\(3 \times 3 \times 3\) and stride 1, followed by batch normalization
and LeakyReLU activation. Besides, in every two convolutional layers is
inserted a max-pooling layer. After convolution, the inputs' shape is
transformed into \(128 \times 5 \times 5 \times 5\). PrimaryCaps layer
is in nature a convolutional layer with its scalar outputs chunked into
vectors and then squashed. The squashing function is the novel
non-linear activation function designed for capsules. DigitCaps layer
receives input from all the capsules in the PrimaryCaps layer with
routing by agreement and produces a matrix with a size of
\(166 \times 8\).

\emph{Decoder.} The decoder part takes the outputs of DigitCaps, the
fingerprint, to reconstruct the input force fields grids. This part is
used as a regularizer, whose job is to encourage the digit capsules to
encode the pose information into each digit. Therefore, we simply use a
nearly symmetric architecture consisting of a linear layer, transposed
3D convolutional layers with upsampling to decode the outputs of
DigitCaps.

\hypertarget{model-training}{%
\subsubsection{Model Training}\label{model-training}}

Our model was implemented with Pytorch 1.3\textsuperscript{46} and
trained with Adam\textsuperscript{47} optimizer. The learning rate was
set to 0.001, and all other parameters were default. The loss function
we used was the same as that of CapsNet, which has two parts: Margin
loss and reconstruction loss.

\[Loss = margin\_ loss + weight*\ recontruction\_\text{loss}\]

where the two decomposed losses use similar forms as Hinton et
al.\textsuperscript{35}:

\[margin\_ loss = \frac{1}{k}\sum_{k}^{}{T_{k}\max{\left( 0,\ m^{+} - \left\| \mathbf{v}_{k} \right\| \right)^{2} + \lambda\left( 1 - T_{k} \right)\max\left( 0,\ \left\| \mathbf{v}_{k} \right\| - m^{-} \right)^{2}}}\]

\[reconstution\_ loss = MSE(inputs,\ reconstructions)\]

where \(k\) is digital capsule index; \(m^{+} = 0.9\) and
\(m^{-} = 0.1\); \(T_{k} = 1\) if a digit of class \(k\) is present;
\(\lambda = 0.5\). The margin loss here was calculated with 2D molecular
fingerprints as labeled digit classes. Both MACCSKey and ECFP4 were
tested here for training. The final model for evaluation was trained on
1\% random samples of the full prepared data set (\textasciitilde{} 3
million compounds, further explained below) with MACCSKey over 10 epochs
with a reconstruction weight of 0.005. All fingerprints were calculated
and stored for training in advance, but the force fields grids were
calculated on the fly. The training process for the final model took up
to about one week on three NVIDIA TITAN RTX GPUs.

\hypertarget{data-set}{%
\subsubsection{Data set}\label{data-set}}

\emph{ZINC15}\textsuperscript{48} 3D lead-like (logP \textless{}= 3.5,
250 \textless{} MW \textless{} 500) subset was retrieved to train our
model. A total number of \textasciitilde{}286 million molecules with
pre-generated conformers were downloaded from the ZINC database.
Approximately 25 thousand molecules (less than 0.01\%) that cannot be
parameterized by the MMFF94 force field or cannot be sanitized by RDKit
(v2019.09.2)\textsuperscript{49, 50} were filtered out and excluded from
the subsequent training process. However, given the huge calculation
cost of force fields grids and the large size of the full data set, not
all these data were actually used in the study. We randomly sampled
various proportions of the full data set and split it randomly into a
training set and test set by 9:1 for model training.

\emph{ChEMBL25}\textsuperscript{51, 52} were retrieved for the
evaluation. The distribution of the number of rotatable bonds was
analyzed for all molecules within ChEMBL25. For the assessment of
fingerprints' sensitivity to 3D structural changes, 1000 molecules were
randomly sampled for each number of rotatable bonds ranging from 0 to
15. To obtain a relatively wide range of RMSD values, 100 conformers
were generated for each molecule, and then 1 conformer was selected as
the reference and another 99 were aligned to the reference with RMSD
calculated and ranked. Finally, 10 out of 99 were selected with equal
RMSD intervals from high to low and calculated similarities to the
reference with various fingerprints. Conformation generation and
alignment were implemented with RDKit (v2019.09.2)\textsuperscript{49,
50}. RMSD was calculated using \emph{rmsdcalc} utility from Schrodinger
Suites 2018-1\textsuperscript{53}.

\emph{PDBbind v2018 General Set}\textsuperscript{54} was retrieved for
the evaluation. Only the complexes with pActivity (pK\textsubscript{d},
pIC\textsubscript{50}, etc.) greater than 6 were kept as ``active''
samples and used subsequently. As a conformation-dependent fingerprint
without rotation invariance, ligand pairs need to be aligned to each
other before fingerprinting and similarity calculation. We investigated
the performance of our model in three situations: i) Query
co-crystallized conformation aligned to co-crystallized conformation
(X2X) with RDKit; ii) Query low-energy conformation aligned to
co-crystallized conformation (LE2X) with RDKit; iii) Query flexibly
aligned to co-crystallized conformation with Omega\textsuperscript{55}
and ROCS\textsuperscript{23}. The first one was aimed to rule out bad
effects resulting from conformation sampling strategy; the latter two
were designed to match the real target prediction scenario.

\emph{The dataset for solubility and malaria bioactivity prediction} was
benchmarked by Duvenaud et al.\textsuperscript{16} and Kearnes et
al.\textsuperscript{17}

\hypertarget{fingerprints-calculation}{%
\subsubsection{Fingerprints
calculation}\label{fingerprints-calculation}}

Except that 166-bits ECFP4 was used only in the model training,
1024-bits one was utilized in both training and all the evaluations.
Both ECFP4 and MACCSKey were computed using RDKit. 1024-bits E3FP was
computed using the default parameters and codes provided by the author
in his GitHub repo
(\href{https://github.com/keiserlab/e3fp-paper/tree/1.1/e3fp_paper}{{https://github.com/keiserlab/e3fp-paper/tree/1.1/e3fp\_paper}}).
USR/USRCAT were also evaluated in this study, calculated with RDKit.

\hypertarget{similarity-calculation-for-two-fingerprints}{%
\subsubsection{Similarity calculation for two
fingerprints}\label{similarity-calculation-for-two-fingerprints}}

Tanimoto coefficient was used for MACCSKey, ECFP4, and E3FP. USRScore
implemented in RDKit was used for USR/USRCAT. As for real-valued TF3P,
since it is a matrix of which each row vector has its independent
structural meaning, it is intuitive to calculate the similarity
row-wisely. We designed a weighted mean of the cosine similarity of each
row vector to calculate the similarity between two TF3Ps,
\(\mathbf{M},\mathbf{\ N \in}\mathbb{R}^{K\mathbf{\times}L}\), as the
following form:

\[\text{Si}m = {(\frac{1}{2k}\sum_{k}^{}{(\left( 1 - \left| m_{k} - n_{k} \right| \right)\frac{p_{k}}{m_{k}n_{k}} + 1)})}^{3}\]

where \(\mathbf{p,m,n} \in \mathbb{R}^{K}\),
\(p_{k} = \sum_{l}^{}{M_{kl}N_{kl}}\),
\(m_{k} = \sqrt{\sum_{l}^{}M_{kl}^{2}}\),
\(n_{k} = \sqrt{\sum_{l}^{}N_{kl}^{2}}\).

\hypertarget{results-discussion}{%
\subsection{Results \& Discussion}\label{results-discussion}}

\hypertarget{training-performance-of-models-with-different-super-parameters}{%
\subsubsection{Training performance of models with different super
parameters
}\label{training-performance-of-models-with-different-super-parameters}}

As each digit represents a certain 2D structural feature, 2D
fingerprints were used as digit class labels during the training
process. Under this circumstance, each digit capsule had to learn to
output an eight-dimensional vector containing the 3D force fields
information in regard to each 2D structural feature.

We first investigated what proportion and epochs for training could make
a compromise between model performance and time cost for training. As
shown in \textbf{Figure 3 A}, the more data was fed into the model, the
lower did the loss value achieve. An interesting point is that the loss
value decreases to almost the same after the same amount of the total
fed data, no matter what proportion of the full dataset sampled. This
indicates a great structural redundancy of the ZINC database. In
summary, 1\% of the full dataset (\textasciitilde{} 3 million compounds)
is sufficient for the convergence of the loss function after 10 epochs
of training with MACCSKey. The grid size (\(\text{GS}\)) and
reconstruction weight (\(W\)) were set to 50 and 0.005, respectively.

\includegraphics[width=6.12222in,height=2.02569in]{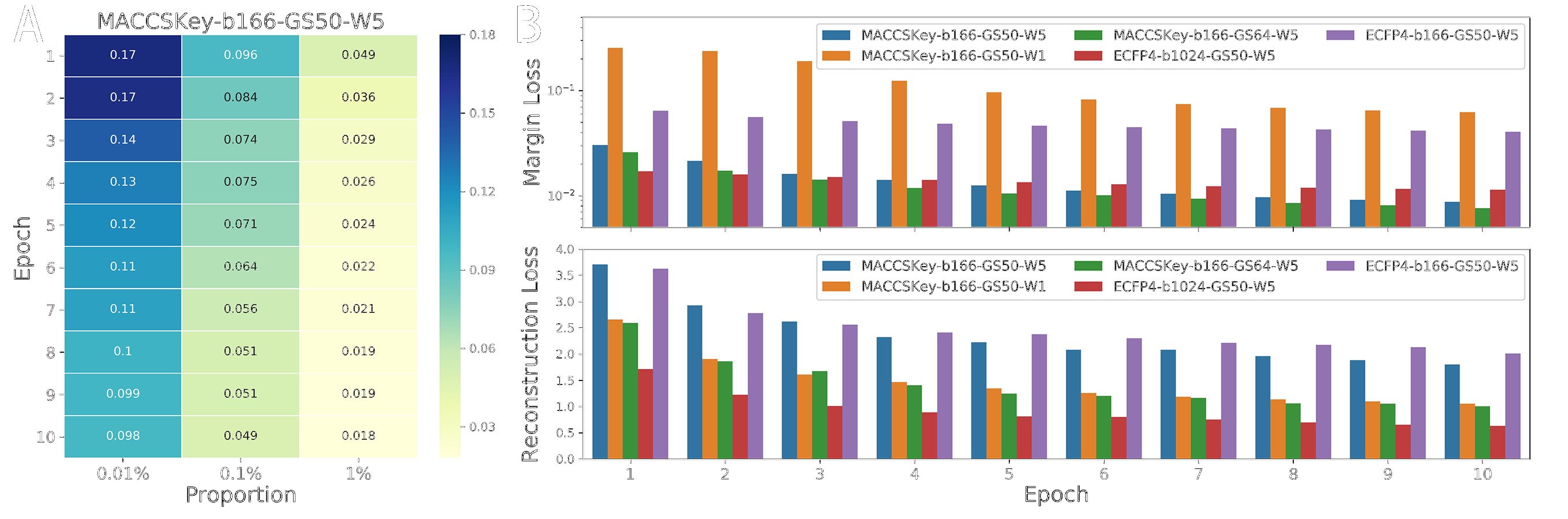}

\textbf{Figure 3.} Training performance with different super parameters.
A) The loss value of test set when training with super parameters as
MACCSKey-b166-GS50-W5. B) The decomposed loss values of test set when
training with indicated super parameters. b166/b1024, 166-bit/1024-bit
version of indicated fingerprint; GS50/GS64, grid size set as
\(50 \times 50 \times 50\)/\(64 \times 64 \times 64\); W1/W5,
reconstruction set as 1/0.005. For loss curves of train/test set during
training, see \textbf{Figure S1}.

After the optimal proportion and training epochs determined, we assessed
how much the training performance dependent on other super parameters of
the model, i.e. \(W\), \(\text{GS}\), and 2D fingerprints. As shown in
\textbf{Figure 3 B}, increasing the \(W\) to 1 would make the
reconstruction loss dominate the margin loss during training
(MACCSKey-b166-GS50-W5 vs. MACCSKey-b166-GS50-W1) but the reconstruction
is expected to play as a regularizer and less important than margin
loss, which was mentioned by Hinton et al.\textsuperscript{35} We next
found that increasing \(\text{GS}\) to \(64 \times 64 \times 64\)
contributed a lot to the decrease of the reconstruction loss but a
little to the margin loss (MACCSKey-b166-GS50-W5 vs.
MACCSKey-b166-GS64-W5). Furthermore, another fingerprint, ECFP4, was
evaluated to show how much different types of fingerprints impacted on
the training performance. To make the results comparable, two versions,
i.e. 166-bit one and 1024-bit one, were included. Compared with
MACCSKey, the loss value convergent pretty slowly when training with
166-bit ECFP4 (MACCSKey-b166-GS50-W5 vs. ECFP4-b166-GS50-W5), but
1024-bit version demonstrated comparable results with the model with
MACCSKey, \(GS = 64\), and \(W = 0.005\) (ECFP4-b1024-GS50-W5 vs.
MACCSKey-b166-GS64-W5).

The capsular networks are often constrained by computation efficiency.
Our models would certainly slow down when the number of network
parameters grew due to either the increase of grid size or the increase
of the number of digital capsules (i.e. the number of bits of 2D
fingerprints). Results from a simple test (\textbf{Table 1}) shows
additional non-trivial time costs when more complex models used, e.g.
MACCSKey-b166-GS64-W5 and ECFP4-b1024-GS50-W5, which retarded further
training with other super parameters, e.g. ECFP4-b1024-GS64-W5. All in
all, three out of five models were retained to conduct the next
evaluation task, namely MACCSKey-b166-GS50-W5, MACCSKey-b166-GS64-W5,
and ECFP4-b1024-GS50-W5, which were referred to as TF3P-50, TF3P-64, and
TF3P-1024, respectively.

\textbf{Table 1.} Time costs of training and inference for models with
different super parameters.

\begin{tabular}{@{}p{80pt}p{60pt}p{60pt}p{60pt}p{60pt}p{60pt}@{}}
\toprule
\multicolumn{4}{c}{Model's Super-parameters} & 
\multirow{2}{60pt}{Training Time\emph{\textsuperscript{a}} (days)} &
\multirow{2}{60pt}{Inference Rates\emph{\textsuperscript{b}} (mols/min)} \tabularnewline
\cline{1-4}
2D Fingerprint & No. of Bits & Grid Size & Reconstruction Weight & & \tabularnewline
\midrule
MACCSKey & 166 & 50 & 0.005 & \textasciitilde{} 7 & \textasciitilde{} 6300 \tabularnewline
MACCSKey & 166 & 50 & 1 & \textasciitilde{} 7 & \textasciitilde{} 6300 \tabularnewline
MACCSKey & 166 & 64 & 0.005 & \textasciitilde{} 10 & \textasciitilde{} 3900 \tabularnewline
ECFP4 & 166 & 50 & 0.005 & \textasciitilde{} 7 & \textasciitilde{} 6300 \tabularnewline
ECFP4 & 1024 & 50 & 0.005 & \textasciitilde{} 10 & \textasciitilde{} 3700 \tabularnewline
\bottomrule
\end{tabular}

\emph{\textsuperscript{a}}: 1\% proportion of the full dataset over 10
epochs on three NVIDIA TITAN RTX GPUs.
\emph{\textsuperscript{b}}: Random samples from ChEMBL on one NVIDIA
GeForce 2080 Ti GPU.

\hypertarget{capturing-3d-structural-changes-of-molecules}{%
\subsubsection{Capturing 3D structural changes of
molecules}\label{capturing-3d-structural-changes-of-molecules}}

The basic virtue of 3D fingerprint is the capability to discriminate
different conformers of a molecule. To assess fingerprints' ability to
capture conformational changes, we sampled 1000 molecules with the
number of rotatable bonds ranging from 0 to 15 (This range was
determined by the distribution of the number of rotatable bonds in
ChEMBL25, \textbf{Figure S2}). The RMSD values and similarity by
fingerprints between different conformers of each molecule were
calculated. As the number of rotatable bonds increases, the overall RMSD
rises, denoting the growing conformational space of molecule
(\textbf{Figure 4}). 2D fingerprints unable to encode 3D conformational
information produce similarity values of one for all entries.
Similarities by three versions of TF3P demonstrated comparably strong
correlations with RMSD, measured by Pearson's r coefficients (-0.71
\textasciitilde{} -0.72) (\textbf{Figure 4} and \textbf{Table S1}).
E3FP, USR, and USRCAT do have the ability to apprehend 3D conformational
changes but yield lower coefficients than TF3P. Taken together, TF3P is
a better predictor, i.e. more sensitive, to the actual 3D structural
changes because TF3P can provide similarity rankings in better
accordance with the 3D structural changes, which is actually we need
when quantifying chemical similarity, despite the fact that TF3P yields
generally higher similarity values. In addition, it is unreasonable that
some pairs of conformers with pretty low RMSD (\textless{} 1.0 Å) yield
low similarities by E3FP (Down left points in \textbf{Figure 4 D}, a
typical example shown in \textbf{Figure S3}).

\includegraphics[width=5.99583in,height=4.56319in]{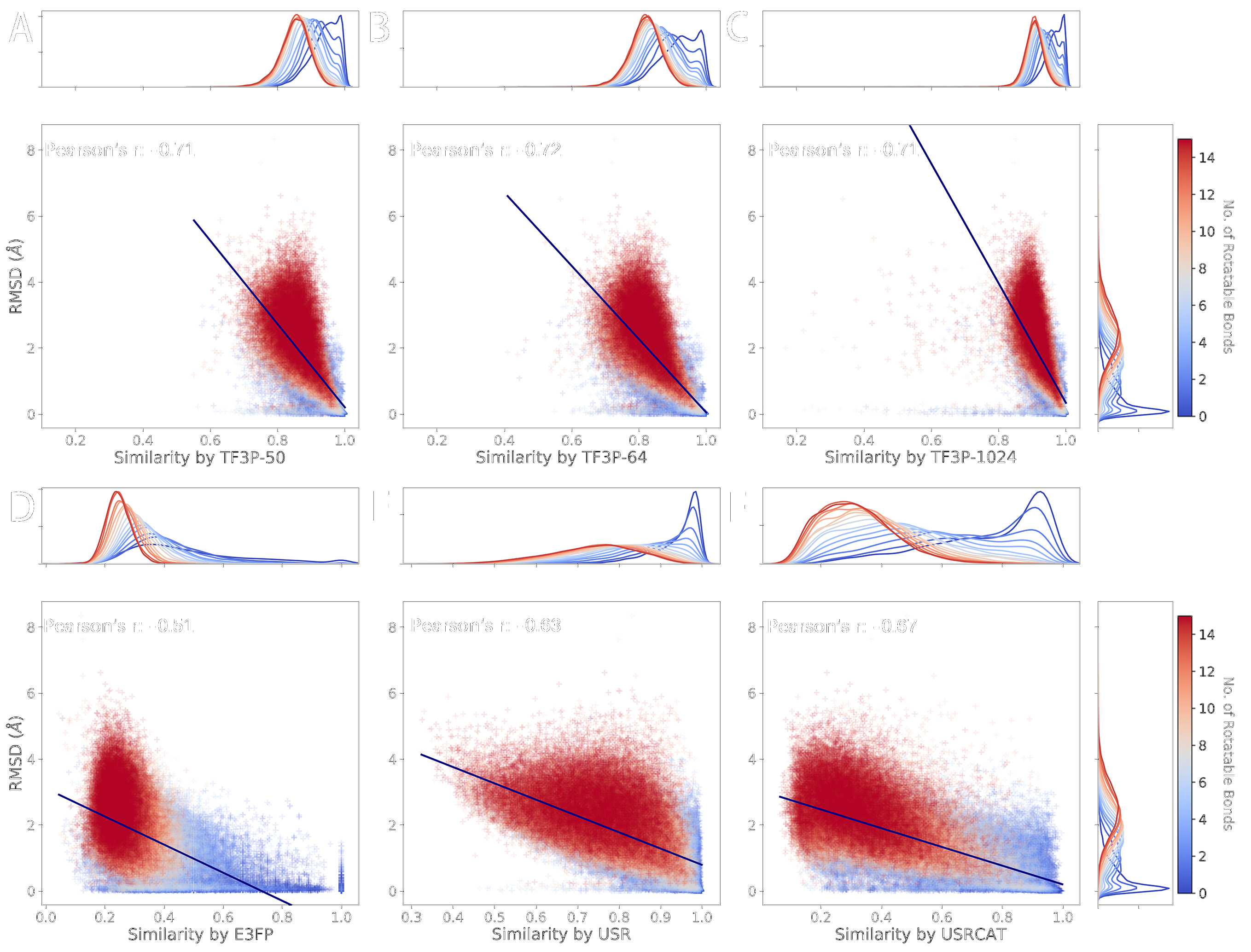}

\textbf{Figure 4.} The correlation between RMSD and similarity by
TF3P-50 (A), TF3P-64 (B), TF3P-1024 (C), E3FP (D), USR (E), and USRCAT
(F). Each point represents a pair of conformers belonging to a molecule,
colored by its number of rotatable bonds. The density plots are colored
in the same way. TF3P-50, TF3P-64, and TF3P-1024 refer to
MACCSKey-b166-GS50-W5, MACCSKey-b166-GS64-W5, and ECFP4-b1024-GS50-W5,
respectively.

Two examples illustrated what discussed above. \textbf{CHEMBL4116653}
can undergo conformational isomerization because of its freely rotatable
single bonds (\textbf{Figure 5 A}). Among three conformers of
\textbf{CHEMBL4116653}, \textbf{conf\_1} is very similar to
\textbf{conf\_3} but far dissimilar to \textbf{conf\_2}, indicated by
the divergence in RMSD values (1.89 Å vs. 6.63 Å). The similarities by
TF3P are consistent with this divergence (0.88 vs. 0.52), but not by
E3FP (0.28 vs. 0.28). Another case is a series of natural products with
four chiral carbons\textsuperscript{56}, a pair of enantiomers plus a
structural isomer (\textbf{Figure 5 B}). 2D fingerprints encounter
bafflement when distinguishing between a pair of enantiomers,
\textbf{Anti-hh} and \textbf{Anti-hh'}, indicated by the similarities of
one. Since they are isomers, the RMSD values among them can be also
calculated to indicate the 3D conformational deviation (2.89 Å vs. 3.81
Å, for \textbf{Anti-hh} \& \textbf{Anti-hh'} and \textbf{Anti-hh} \&
\textbf{Anti-ht}, respectively). Similar situations occur for TF3P and
E3FP, respectively: The similarity by E3FP hardly changes (0.30 vs.
0.29), whereas the similarity by TF3P shifts accordingly with the RMSD
(0.78 vs. 0.73). Besides, there are also counter examples where TF3P
failed to capture some important conformational changes. As shown in
\textbf{Figure 5 C}, \textbf{CHEMBL1789181} undergoes conformational
isomerization with part of the structure flipped over, the aromatic ring
moiety and the aliphatic chain moiety swapping position with each other.
Unlike USRCAT and E3FP, TF3P showed its deficiency in apprehending this
kind of change.

\includegraphics[width=5.99583in,height=2.45139in]{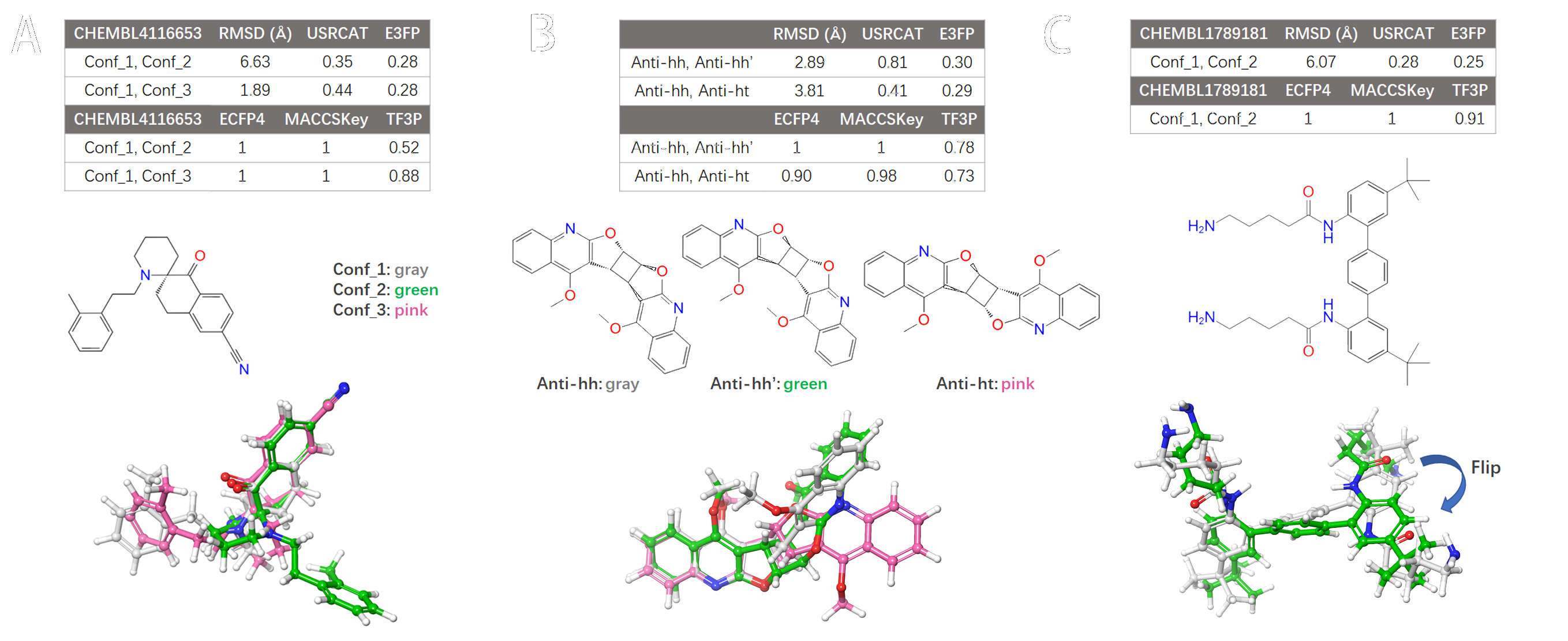}

\textbf{Figure 5.} A, B) Two promising examples with RMSD, similarities,
2D structures and 3D aligned conformers. C) One counter example with
RMSD, similarities, 2D structures and 3D aligned conformers. TF3P here
refers to TF3P-50.

\hypertarget{finding-similar-pockets-by-similar-ligands}{%
\subsubsection{Finding similar pockets by similar
ligands}\label{finding-similar-pockets-by-similar-ligands}}

The primary intuition of using molecular 3D force fields grids has two
aspects. On the one hand, force fields contain the raw information
dominating the interaction between drugs and targets; on the other hand,
force fields are degenerate of atom types and bonds of molecular graph
representation. TF3P, as a fingerprint derived from force fields,
therefore, was expected to be capable to recognize molecule pairs that
are dissimilar in 2D topology but resemble each other in 3D shape and
electrostatics. Given the 3D complementation between the ligands and the
pockets they bind to, this is deemed to improve the precision of the
targets similarity inference based on ligands similarity, which is the
essence of similarity-based target prediction. In this study, targets
similarity was represented by FuzCav pockets
similarity\textsuperscript{57}, with particular emphasis on the binding.
PDBbind v2018 General Set was selected to study the inference further
with three situations (i.e. X2X, LE2X, and Flex) evaluated to meet the
concerns on conformation generation, as mentioned previously in Methods
\& Materials. Given the time costs and performances discussed before, we
only included TF3P-50 out of three models in this part and referred to
as simply as TF3P.

As shown in \textbf{Table 2}, the overall (Top 100\%) Pearson's r
coefficients of FuzCav similarity vs. the similarity by all fingerprints
in all situations are low (0.08-0.20), suggesting the toughness to
perfectly infer target similarity from only ligands information.
Nevertheless, it is noteworthy that the correlations are generally
higher in top-ranked samples by similarity than bottom-ranked ones with
no exception. The same trend occurs in Spearman's rho (\textbf{Table 3})
and Kendall's tau coefficients (\textbf{Table 4}). This probably results
from that subtle differences between similar ligands correspond to a
similar degree of conformational changes between pockets and thus
presenting correlations easy to fit for top-ranked pairs by fingerprints
similarity. However, in contrast to similar ligands binding to pockets
alike, every ligand dissimilar to others binds to its target in its own
way, causing complications to the regression for dissimilar pairs. As
for top-ranked pairs, good correlations only mean similar orders in two
arrays of samples, not the exact values. We further analyzed the average
values of FuzCav similarities of samples with incrementing ranking
thresholds. As shown in \textbf{Figure 6}, top-ranked pairs by TF3P
similarity present significantly higher average FuzCav pocket
similarities than those by USRCAT, E3FP and 2D fingerprints under the
same situation. Taken together, compared to other fingerprints, similar
ligands found by TF3P do bind to similar targets and their ligand
similarity is a better indicator of the pocket similarity.

\textbf{Table 2.} Pearson correlation coefficients of FuzCav similarity
vs. different fingerprints similarity for all active pairs from PDBbind
v2018 General Set.

\begin{tabular}{@{}p{20pt}p{20pt}p{20pt}p{20pt}p{1pt}p{20pt}p{20pt}p{20pt}p{1pt}p{20pt}p{20pt}p{20pt}p{25pt}p{25pt}p{30pt}@{}}
\toprule
\multirow{2}{20pt}{Top \%\emph{\textsuperscript{a}}} &
\multicolumn{3}{c}{X2X\emph{\textsuperscript{b}}} &
\multirow{2}{1pt}{} &
\multicolumn{3}{c}{LE2X\emph{\textsuperscript{c}}} &
\multirow{2}{1pt}{} &
\multicolumn{4}{c}{Flex\emph{\textsuperscript{d}}} &
\multirow{2}{25pt}{MACCS Key} &
\multirow{2}{30pt}{ECFP4}
\tabularnewline
\cline{2-4}
\cline{6-8}
\cline{10-13}
& TF3P & E3FP & USR CAT & & TF3P & E3FP & USR CAT & & TF3P & E3FP & USR CAT
& ROCS-Combo & &
\tabularnewline
\midrule
0.5 & \textbf{0.52*} & 0.29 & 0.40 & & 0.24 & 0.19 & 0.22 & & 0.36 &
0.24 & 0.30 & 0.37 & 0.34 & 0.31\tabularnewline
1 & \textbf{0.47*} & 0.30 & 0.36 & & 0.22 & 0.21 & 0.19 & & 0.32 & 0.26
& 0.27 & 0.37 & 0.32 & 0.29\tabularnewline
5 & 0.24 & 0.25 & 0.25 & & 0.12 & 0.20 & 0.13 & & 0.15 & 0.23 & 0.18 &
\textbf{0.27*} & 0.20 & 0.24\tabularnewline
10 & 0.17 & 0.22 & 0.21 & & 0.08 & 0.17 & 0.11 & & 0.10 & 0.21 & 0.16 &
\textbf{0.23*} & 0.14 & 0.21\tabularnewline
20 & 0.12 & \textbf{0.19*} & 0.18 & & 0.06 & 0.15 & 0.11 & & 0.08 & 0.18
& 0.14 & 0.18 & 0.10 & 0.18\tabularnewline
50 & 0.07 & 0.15 & 0.15 & & 0.06 & 0.13 & 0.11 & & 0.06 & 0.15 & 0.13 &
\textbf{0.17*} & 0.08 & 0.14\tabularnewline
100 & 0.08 & 0.13 & 0.17 & & 0.08 & 0.12 & 0.15 & & 0.09 & 0.13 & 0.15 &
\textbf{0.20*} & 0.10 & 0.11\tabularnewline
\bottomrule
\end{tabular}

\textsuperscript{a}: Sorted by the indicated fingerprint similarity.
\textsuperscript{b}: Co-crystallized conformation aligned to
co-crystallized conformation. \textsuperscript{c}: Low-energy
conformation aligned to co-crystallized conformation.
\textsuperscript{d}: Flexibly aligned to co-crystallized conformation.

\textbf{Table 3.} Spearman rank correlation coefficients of FuzCav
similarity vs. different fingerprints similarity for all active pairs
from PDBbind v2018 General Set.

\begin{tabular}{@{}p{20pt}p{20pt}p{20pt}p{20pt}p{1pt}p{20pt}p{20pt}p{20pt}p{1pt}p{20pt}p{20pt}p{20pt}p{25pt}p{25pt}p{30pt}@{}}
\toprule
\multirow{2}{20pt}{Top \%\emph{\textsuperscript{a}}} &
\multicolumn{3}{c}{X2X\emph{\textsuperscript{b}}} &
\multirow{2}{1pt}{} &
\multicolumn{3}{c}{LE2X\emph{\textsuperscript{c}}} &
\multirow{2}{1pt}{} &
\multicolumn{4}{c}{Flex\emph{\textsuperscript{d}}} &
\multirow{2}{25pt}{MACCS Key} &
\multirow{2}{30pt}{ECFP4}
\tabularnewline
\cline{2-4}
\cline{6-8}
\cline{10-13}
& TF3P & E3FP & USR CAT & & TF3P & E3FP & USR CAT & & TF3P & E3FP & USR CAT
& ROCS-Combo & &
\tabularnewline
\midrule
0.5 & \textbf{0.33*} & 0.16 & 0.13 & & 0.14 & 0.12 & 0.06 & & 0.21 &
0.15 & 0.10 & 0.20 & 0.18 & 0.19\tabularnewline
1 & \textbf{0.20*} & 0.12 & 0.08 & & 0.08 & 0.10 & 0.04 & & 0.12 & 0.11
& 0.06 & 0.14 & 0.12 & 0.11\tabularnewline
5 & -0.01 & \textbf{0.08*} & 0.07 & & -0.04 & 0.07 & 0.05 & & -0.04 &
\textbf{0.08*} & 0.06 & 0.06 & 0.01 & 0.03\tabularnewline
10 & -0.03 & 0.07 & \textbf{0.08*} & & -0.05 & 0.07 & 0.05 & & -0.05 &
\textbf{0.08*} & 0.07 & 0.05 & 0.00 & 0.02\tabularnewline
20 & -0.03 & 0.07 & \textbf{0.08*} & & -0.03 & 0.07 & 0.06 & & -0.03 &
\textbf{0.08*} & 0.07 & 0.06 & 0.01 & 0.01\tabularnewline
50 & -0.00 & 0.07 & 0.09 & & 0.00 & 0.07 & 0.07 & & 0.01 & 0.08 & 0.08 &
\textbf{0.11*} & 0.03 & 0.02\tabularnewline
100 & 0.04 & 0.08 & 0.13 & & 0.05 & 0.08 & 0.13 & & 0.01 & 0.09 & 0.12 &
\textbf{0.20*} & 0.08 & 0.04\tabularnewline
\bottomrule
\end{tabular}

\textsuperscript{a}: Sorted by the indicated fingerprint similarity.
\textsuperscript{b}: Co-crystallized conformation aligned to
co-crystallized conformation. \textsuperscript{c}: Low-energy
conformation aligned to co-crystallized conformation.
\textsuperscript{d}: Flexibly aligned to co-crystallized conformation.

\textbf{Table 4.} Kendall rank correlation coefficients of FuzCav
similarity vs. different fingerprints similarity for all active pairs
from PDBbind v2018 General Set.

\begin{tabular}{@{}p{20pt}p{20pt}p{20pt}p{20pt}p{1pt}p{20pt}p{20pt}p{20pt}p{1pt}p{20pt}p{20pt}p{20pt}p{25pt}p{25pt}p{30pt}@{}}
\toprule
\multirow{2}{20pt}{Top \%\emph{\textsuperscript{a}}} &
\multicolumn{3}{c}{X2X\emph{\textsuperscript{b}}} &
\multirow{2}{1pt}{} &
\multicolumn{3}{c}{LE2X\emph{\textsuperscript{c}}} &
\multirow{2}{1pt}{} &
\multicolumn{4}{c}{Flex\emph{\textsuperscript{d}}} &
\multirow{2}{25pt}{MACCS Key} &
\multirow{2}{30pt}{ECFP4}
\tabularnewline
\cline{2-4}
\cline{6-8}
\cline{10-13}
& TF3P & E3FP & USR CAT & & TF3P & E3FP & USR CAT & & TF3P & E3FP & USR CAT
& ROCS-Combo & &
\tabularnewline
\midrule
0.5 & \textbf{0.22*} & 0.11 & 0.08 & & 0.09 & 0.08 & 0.04 & & 0.14 &
0.10 & 0.07 & 0.13 & 0.12 & 0.12\tabularnewline
1 & \textbf{0.13*} & 0.08 & 0.05 & & 0.05 & 0.07 & 0.02 & & 0.08 & 0.08
& 0.04 & 0.09 & 0.08 & 0.08\tabularnewline
5 & -0.01 & \textbf{0.05*} & 0.04 & & -0.03 & \textbf{0.05*} & 0.03 & &
-0.03 & \textbf{0.05*} & 0.04 & 0.04 & 0.01 & 0.02\tabularnewline
10 & -0.02 & \textbf{0.05*} & \textbf{0.05*} & & -0.03 & 0.04 & 0.04 & &
-0.03 & \textbf{0.05*} & \textbf{0.05*} & 0.03 & 0.00 &
0.01\tabularnewline
20 & -0.02 & 0.05 & \textbf{0.06*} & & -0.02 & 0.04 & 0.04 & & -0.02 &
0.05 & 0.05 & 0.04 & 0.01 & 0.01\tabularnewline
50 & -0.00 & 0.05 & 0.06 & & 0.00 & 0.04 & 0.05 & & 0.01 & 0.05 & 0.05 &
\textbf{0.08*} & 0.02 & 0.01\tabularnewline
100 & 0.03 & 0.05 & 0.09 & & 0.04 & 0.05 & 0.09 & & 0.04 & 0.06 & 0.08 &
\textbf{0.13*} & 0.05 & 0.03\tabularnewline
\bottomrule
\end{tabular}

\textsuperscript{a}: Sorted by the indicated fingerprint similarity.
\textsuperscript{b}: Co-crystallized conformation aligned to
co-crystallized conformation. \textsuperscript{c}: Low-energy
conformation aligned to co-crystallized conformation.
\textsuperscript{d}: Flexibly aligned to co-crystallized conformation.

\includegraphics[width=5.99306in,height=2.52639in]{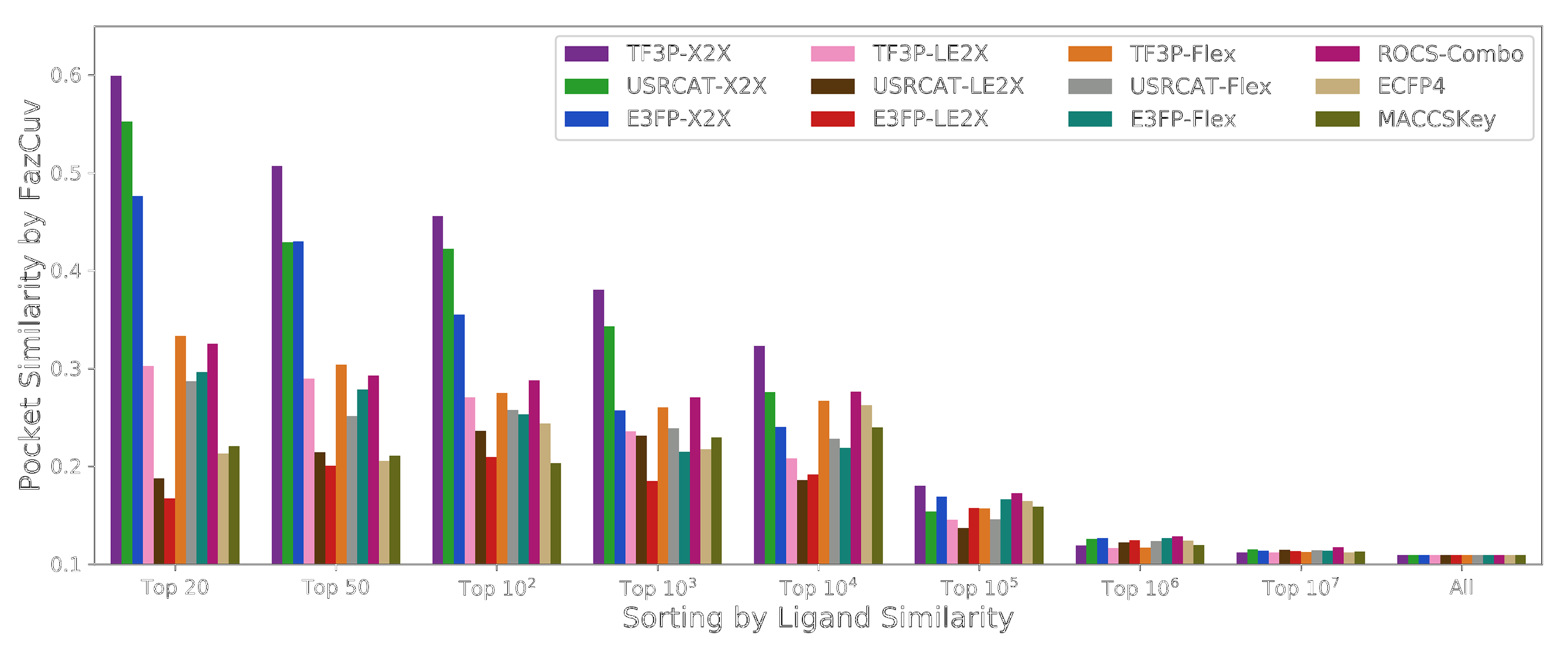}

\textbf{Figure 6.} Average pocket similarity for samples with
incrementing cutoffs in ranking (Top 1\% is equivalent to top
10\textsuperscript{5}).

In fact, we only pay attention to top-ranked outputs in the real
scenarios of ligand-based drug discovery. Like the early-stage
enrichment is of vital importance for virtual screening, we also need a
stronger ``early-stage enrichment'' for target prediction based on only
ligands similarity. This is to say, a good fingerprint for ligand-based
target prediction need have the ability to rank the targets whose
pockets are similar to the true target in the top of the targets pool
and with an order that is well correlated to their pockets similarity.
To this end, TF3P outperforms the existing 3D and 2D fingerprints,
indicated by the higher average FuzCav similarities and higher
correlations coefficients under all three situations in top 1\% shown in
\textbf{Table 2, 3,} and \textbf{4,} and \textbf{Figure 6}. This is not
that mention that TF3P-LE2X can achieve comparable results with all
fingerprints under Flex situation (\textbf{Figure 6}).

Additionally, another noteworthy point is that the correlation for top
ranked ones can be generally ranked as: X2X \textgreater{} Flex
\textgreater{} LE2X. This is of great significance because many reported
that 2D fingerprints outperformed 3D descriptors under the condition of
using generated low-energy conformers\textsuperscript{58, 59}. Our
results suggest that those barely satisfactory performance of 3D
fingerprints are likely to be attributed to not use co-crystallized
conformer (i.e. bioactive conformer), which urges more endeavors to
develop new computational tools meeting this demand.

Here, we also show three promising examples among the most similar pairs
by TF3P that are not alike by other fingerprints (\textbf{Figure 7 A,
B,} and \textbf{C}), and one counter example (\textbf{Figure 7 D}). Each
pair binds to the same target. As the similarity values calculated by
different fingerprints over the same dataset show divergent
distributions (\textbf{Figure S4}), the percentile ranks were computed
and attached below the exact values of similarity to demonstrate how
similar different fingerprints regard them. As shown in the aligned
crystal structures (\textbf{Figure 7 A, B,} and \textbf{C}), all three
pairs have obviously similar 3D force fields but dissimilar in 2D
topology. This leads to the difficulty to recognize their resemblance
for the existing fingerprints based on topological structure but not for
TF3P that is derived from 3D force fields. Therefore, T3FP can rank
these pairs at the top places (Top 1\%) based on ligands similarity but
others cannot. However, TF3P can miss under some circumstances.
\textbf{Figure 7 D} shows a typical case where two ligands bind to the
same target but have distinct structures. Since the multi-benzene moiety
and the aliphatic chain both serve as spacer, only the shape of the
spacer counts here but not the charges, providing a possible explanation
for the behavior of TF3P that is based on both vdW potential and
electrostatic potential.

\includegraphics[width=5.99306in,height=4.22083in]{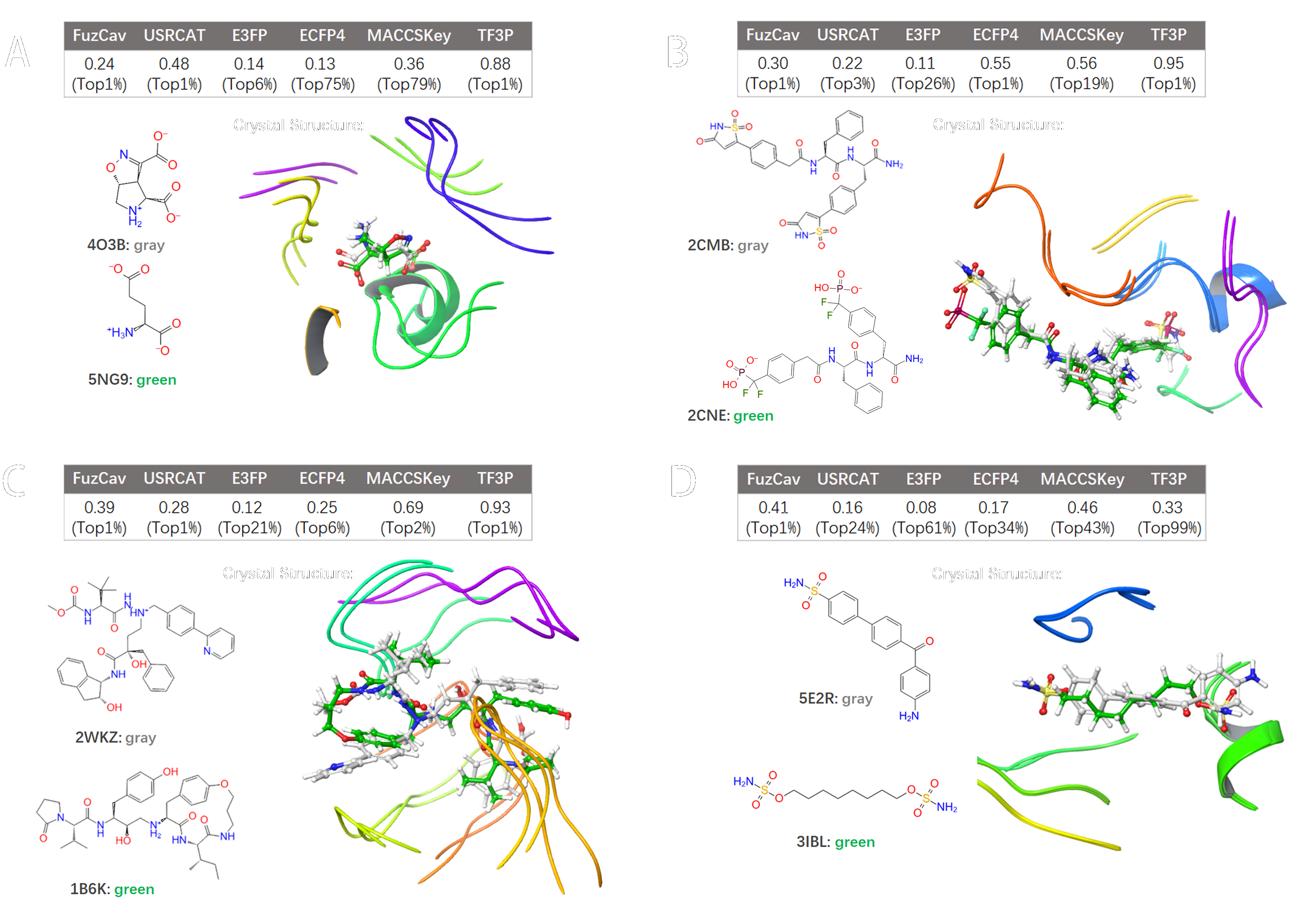}

\textbf{Figure 7.} A, B, C) Three top-ranked similar pairs by TF3P but
are inaccessible by other fingerprints. In terms of receptors, GluA2 for
4O3B and 5NG9, Protein Tyrosine Phosphatase 1B for 2CMB and 2CNE, HIV-1
Protease for 2WKZ and 1B6K. D) One counter example TF3P yielded
divergent results against FuzCav. Human carbonic anhydrase II for 5E2R
and 3IBL.

\hypertarget{sea-analysis-of-pdbbind-database}{%
\subsubsection{SEA analysis of PDBbind
database}\label{sea-analysis-of-pdbbind-database}}

SEA is a widely used statistical model that quantitatively groups and
relates proteins based on the chemical similarity of their ligands. We
utilized SEA to analyze all targets that annotated no less than 10
active complexes within PDBbind v2018 General Set with five
fingerprints, namely TF3P, USRCAT, E3FP, MACCSKey and ECFP4. The results
demonstrate a similar trend as the above study: TF3P can enrich more
pairs of the targets that resemble each other by pocket similarity in
the targets pool (\textbf{Figure 8 A}), yielding higher average FuzCav
similarity in top-ranked pairs. This indicates TF3P as a better choice
for SEA-based target prediction, compared to other fingerprints.

One primary application of SEA is to cluster proteins using only ligands
similarity to find some links unexpected. First, it is not beyond
anticipation that TF3P outputs a clustering of targets closer to FuzCav
similarity, measured by adjusted Rand score (\textbf{Figure 8 B}; for
specific explanation of adjusted Rand score, see Supplementary File 1).
Moreover, some interesting links indeed emerged when we conducted SEA
analysis with TF3P (\textbf{Figure 8 C}). The first one is a cluster of
HIV relevant proteins, including HIV protease, HIV integrase, and HIV
reverse transcriptase, which are highly pharmacologically related. The
second one is endothiapepsin (Uniprot ID: P11838) from
\emph{Cryphonectria parasitica} and plasmepsin-2 (Uniprot ID: P46925)
from \emph{Plasmodium falciparum}, which function similarly as
aspartic-type endopeptidase although they are from different organisms.
The third one is a pair of nuclear receptors, retinoic acid receptor
RXR-alpha (Uniprot ID: P19793) and peroxisome proliferator-activated
receptor gamma (Uniprot ID: P37231), which are not only biologically
related but also have a very conserved 3D structure of the ligand
binding domain (\textbf{Figure 8 D}). Notably, the latter two did not
emerge when using other four fingerprints (\textbf{Figure S5}).

\includegraphics[width=5.99861in,height=4.95625in]{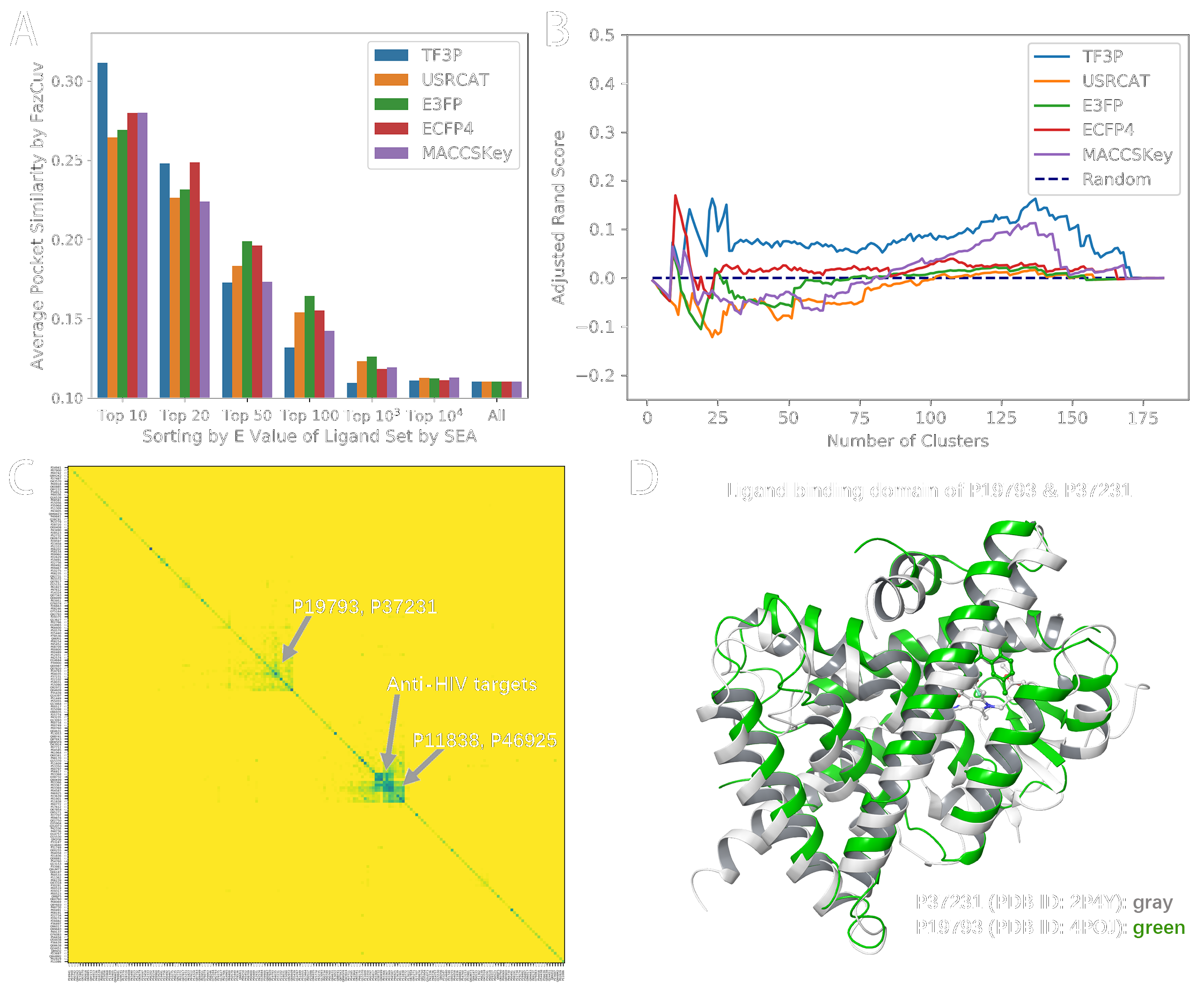}

\textbf{Figure 8.} SEA analysis of the PDBbind database. A) Average
pocket similarity for targets pairs with incrementing cutoffs in the
ranking. B) The distance of FuzCav clustering of targets to clustering
by five fingerprints, measured by adjusted Rand score. Higher means
closer. C) Heatmap of E-value of indicated targets calculated by SEA.
Each point is colored by its E-value with lower values as green and
higher as yellow. For high-resolution images, see Supplementary File 2.
D) Aligned crystal structures of ligand binding domains of two indicated
targets.

\hypertarget{application-with-machine-learning-models}{%
\subsubsection{Application with machine learning
models}\label{application-with-machine-learning-models}}

Aside from calculating the similarity between two molecules, another
important application of fingerprints is to be used as the inputs for
machine learning models. To this end, we used a very simple model,
single linear layer neural network (i.e. linear regression), to
demonstrate TF3P's compatibility and performance on several tasks. TF3P
achieves the lowest MSE value in 5-fold cross-validation among all of
the five fingerprints on solubility prediction and outperforms other 3D
fingerprints on malaria bioactivity prediction (\textbf{Table 5}),
making itself a promising alternative when integrated with machine
learning models.

\textbf{Table 5.} Prediction of solubility and malaria bioactivity with
different fingerprints.

\begin{longtable}[]{@{}lll@{}}
\toprule
\begin{minipage}[b]{0.30\columnwidth}\raggedright
Fingerprint\strut
\end{minipage} & \begin{minipage}[b]{0.30\columnwidth}\raggedright
Solubility \textsuperscript{a}

log10 (mol/L)\strut
\end{minipage} & \begin{minipage}[b]{0.30\columnwidth}\raggedright
Malaria bioactivity \textsuperscript{a}

ln ($\mu$mol/L)\strut
\end{minipage}\tabularnewline
\midrule
\endhead
MACCSKey + Linear layer & 1.33 ± 0.20 & 1.22 ± 0.03\tabularnewline
ECFP4 + Linear layer & 1.71 ± 0.27 & \textbf{1.14 ±
0.03*}\tabularnewline
E3FP + Linear layer & 1.71 ± 0.27 & 1.34 ± 0.06\tabularnewline
USRCAT + Linear layer & 1.08 ± 0.14 & 1.40 ± 0.03\tabularnewline
TF3P + Linear layer & \textbf{0.66 ± 0.09*} & 1.27 ± 0.08\tabularnewline
\bottomrule
\end{longtable}

\textsuperscript{a}: Evaluation matrix: mean ± std. of MSE

\hypertarget{conclusion}{%
\subsection{Conclusion}\label{conclusion}}

In summary, we developed a new fingerprint of 3D molecule with the deep
capsular network, which learned to encode the 3D force fields
information into each digit of 2D fingerprint without labeled data for
specific predictive tasks. Ligands similarity can be intuitively
calculated with the fingerprint produced by our model, TF3P, and thus
making it compatible with statistical models. Also, TF3P could be
applied with machine learning models as the inputs and present promising
results. Since TF3P is derived from molecular 3D force fields, it
demonstrated its ability to recognize molecule pairs that are dissimilar
in 2D topology but resemble each other in 3D shape and electrostatics,
which improves its ability to infer pockets similarity based on only
ligands similarity, detecting pairs of ligands with similar targets
inaccessible by other fingerprints. TF3P is anticipated to be a better
choice of fingerprints for ligand-based drug discovery in the future.
Further development and benchmarking studies of TF3P-based target
prediction software are still going on.

\hypertarget{associated-content}{%
\subsection{Associated Content}\label{associated-content}}

\hypertarget{supplementary-information}{%
\subsubsection{Supplementary
Information}\label{supplementary-information}}

Supplementary File 1: Supplementary figures and tables.

Supplementary File 2: Supplementary files.

\hypertarget{acknowledgments}{%
\subsection{Acknowledgments}\label{acknowledgments}}

This research was supported by the National Key Research and Development
Project (Grant numbers 2019YFC1708900), the National Natural Science
Foundation of China (Grant numbers 81872730, 81673279, 21772005), the
National Major Scientific and Technological Special Project for
Significant New Drugs Development (2019ZX09204-001, 2018ZX09735001-003)
and the Beijing Natural Science Foundation (7202088, 7172118).

\hypertarget{abbreviations-used}{%
\subsection{Abbreviations used}\label{abbreviations-used}}

ECFP, Extended Connectivity Fingerprint; E3FP, Extended 3-Dimensional
Fingerprint; DNN, Deep Neural Network; SEA, Similarity Ensemble
Approach; ROCS, Rapid Overlay of Chemical Structures; QSAR, Quantitative
Structure-Activity Relationship; MMFF, Merck Molecular Force Field;
ReLU, Rectified Linear Unit; ZINC, ZINC Is Not Commercial; MSE, Mean
Squared Error; RMSD, Root-Mean-Square Deviation of Atomic Positions,
IUPAC, International Union of Pure and Applied Chemistry; SMILES,
Simplified Molecular-Input Line-Entry System.

\hypertarget{declaration-of-interest}{%
\subsection{Declaration of interest}\label{declaration-of-interest}}

The authors report no conflicts of interest. The authors alone are
responsible for the content and writing of this article.

\hypertarget{references}{%
\subsection{{References}\label{references}}}

{1.{~~~ }Boström, J.; Hogner, A.; Schmitt,
S., Do Structurally Similar Ligands Bind in a Similar Fashion? \emph{J.
Med. Chem.} \textbf{2006}, 49, 6716-6725.}

{2.{~~~ }Karelson, M.; Lobanov, V. S.;
Katritzky, A. R., Quantum-Chemical Descriptors in Qsar/Qspr Studies.
\emph{Chem. Rev.} \textbf{1996}, 96, 1027-1044.}

{3.{~~~ }Todeschini, R.; Consonni, V.,
\emph{Handbook of Molecular Descriptors. Wileyvch, Weinheim}. 2000; Vol.
11.}

{4.{~~~ }Jiang, P.; Saydam, S.; Ramandi,
H. L.; Crosky, A.; Maghrebi, M. Deep Molecular Representation in
Cheminformatics. In \emph{Handbook of Deep Learning Applications},
Balas, V. E.; Roy, S. S.; Sharma, D.; Samui, P., Eds.; Springer
International Publishing: Cham, 2019, pp 147-159.}

{5.{~~~ }Li, X.; Li, Z.; Wu, X.; Xiong,
Z.; Yang, T.; Fu, Z.; Liu, X.; Tan, X.; Zhong, F.; Wan, X.; Wang, D.;
Ding, X.; Yang, R.; Hou, H.; Li, C.; Liu, H.; Chen, K.; Jiang, H.;
Zheng, M., Deep Learning Enhancing Kinome-Wide Polypharmacology
Profiling: Model Construction and Experiment Validation. \emph{J. Med.
Chem.} \textbf{2019}, In press.}

{6.{~~~ }Durant, J. L.; Leland, B. A.;
Henry, D. R.; Nourse, J. G., Reoptimization of Mdl Keys for Use in Drug
Discovery. \emph{J. Chem. Inf. Comput. Sci.} \textbf{2002}, 42,
1273-1280.}

{7.{~~~ }Rogers, D.; Hahn, M.,
Extended-Connectivity Fingerprints. \emph{J. Chem. Inf. Model.}
\textbf{2010}, 50, 742-754.}

{8.{~~~ }Carhart, R. E.; Smith, D. H.;
Venkataraghavan, R., Atom Pairs as Molecular Features in
Structure-Activity Studies: Definition and Applications. \emph{J. Chem.
Inf. Comput. Sci.} \textbf{1985}, 25, 64-73.}

{9.{~~~ }Nilakantan, R.; Bauman, N.;
Dixon, J. S.; Venkataraghavan, R., Topological Torsion: A New Molecular
Descriptor for Sar Applications. Comparison with Other Descriptors.
\emph{J. Chem. Inf. Comput. Sci.} \textbf{1987}, 27, 82-85.}

{10.{~ }Gedeck, P.; Rohde, B.; Bartels,
C., Qsar - How Good Is It in Practice? Comparison of Descriptor Sets on
an Unbiased Cross Section of Corporate Data Sets. \emph{J. Chem. Inf.
Model.} \textbf{2006}, 46, 1924-1936.}

{11.{~ }Huo, H.; Rupp, M., Unified
Representation of Molecules and Crystals for Machine Learning.
\emph{arXiv preprint arXiv:1704.06439} \textbf{2017}.}

{12.{~ }Sch tt, K. T.; Arbabzadah, F.;
Chmiela, S.; M ller, K. R.; Tkatchenko, A., Quantum-Chemical Insights
from Deep Tensor Neural Networks. \emph{Nat. Commun.} \textbf{2017}, 8,
13890.}

{13.{~ }Lubbers, N.; Smith, J. S.; Barros,
K., Hierarchical Modeling of Molecular Energies Using a Deep Neural
Network. \emph{J. Chem. Phys.} \textbf{2018}, 148, 241715.}

{14.{~ }Sch tt, K. T.; Sauceda, H. E.;
Kindermans, P. J.; Tkatchenko, A.; M ller, K. R., Schnet C a Deep
Learning Architecture for Molecules and Materials. \emph{J. Chem. Phys.}
\textbf{2018}, 148, 241722.}

{15.{~ }Xiong, Z.; Wang, D.; Liu, X.;
Zhong, F.; Wan, X.; Li, X.; Li, Z.; Luo, X.; Chen, K.; Jiang, H.; Zheng,
M., Pushing the Boundaries of Molecular Representation for Drug
Discovery with the Graph Attention Mechanism. \emph{J. Med. Chem.}
\textbf{2019}, In press.}

{16.{~ }Duvenaud, D. K.; Maclaurin, D.;
Iparraguirre, J.; Bombarell, R.; Hirzel, T.; Aspuru-Guzik, A.; Adams, R.
P. Convolutional Networks on Graphs for Learning Molecular Fingerprints.
In Advances in Neural Information Processing Systems, 2015; 2015; pp
2224-2232.}

{17.{~ }Kearnes, S.; McCloskey, K.;
Berndl, M.; Pande, V.; Riley, P., Molecular Graph Convolutions: Moving
Beyond Fingerprints. \emph{J. Comput. Aided Mol. Des.} \textbf{2016},
30, 595-608.}

{18.{ }Shin, W.-H.; Zhu, X.; Bures, G. M.;
Kihara, D., Three-Dimensional Compound Comparison Methods and Their
Application in Drug Discovery. \emph{Molecules} \textbf{2015}, 20,
12841-12862.}

{19.{~ }Schreyer, A. M.; Blundell, T.,
Usrcat: Real-Time Ultrafast Shape Recognition with Pharmacophoric
Constraints. \emph{J. Cheminform.} \textbf{2012}, 4, 27.}

{20.{~ }Ballester, P. J.; Richards, W. G.,
Ultrafast Shape Recognition to Search Compound Databases for Similar
Molecular Shapes. \emph{J. Comput. Chem.} \textbf{2007}, 28, 1711-1723.}

{21.{~ }Axen, S. D.; Huang, X.-P.; C
ceres, E. L.; Gendelev, L.; Roth, B. L.; Keiser, M. J., A Simple
Representation of Three-Dimensional Molecular Structure. \emph{J. Med.
Chem.} \textbf{2017}, 60, 7393-7409.}

{22.{~ }Keiser, M. J.; Roth, B. L.;
Armbruster, B. N.; Ernsberger, P.; Irwin, J. J.; Shoichet, B. K.,
Relating Protein Pharmacology by Ligand Chemistry. \emph{Nat.
Biotechnol.} \textbf{2007}, 25, 197-206.}

{23.{~ }Hawkins, P. C. D.; Skillman, A.
G.; Nicholls, A., Comparison of Shape-Matching and Docking as Virtual
Screening Tools. \emph{J. Med. Chem.} \textbf{2007}, 50, 74-82.}

{24.{~ }Cai, C.; Gong, J.; Liu, X.; Gao,
D.; Li, H., Simg: An Alignment Based Method for Evaluating the
Similarity of Small Molecules and Binding Sites. \emph{J. Chem. Inf.
Model.} \textbf{2013}, 53, 2103-2115.}

{25.{~ }Fontaine, F.; Bolton, E.;
Borodina, Y.; Bryant, S. H., Fast 3d Shape Screening of Large Chemical
Databases through Alignment-Recycling. \emph{Chem. Cent. J.}
\textbf{2007}, 1, 12.}

{26.{~ }Koes, D. R.; Camacho, C. J.,
Shape-Based Virtual Screening with Volumetric Aligned Molecular Shapes.
\emph{J. Comput. Chem.} \textbf{2014}, 35, 1824-1834.}

{27.{ }Wallach, I.; Dzamba, M.; Heifets,
A., Atomnet: A Deep Convolutional Neural Network for Bioactivity
Prediction in Structure-Based Drug Discovery. \emph{arXiv preprint
arXiv:1510.02855} \textbf{2015}.}

{28.{~ }Gomes, J.; Ramsundar, B.;
Feinberg, E. N.; Pande, V. S., Atomic Convolutional Networks for
Predicting Protein-Ligand Binding Affinity. \emph{arXiv preprint
arXiv:1703.10603} \textbf{2017}.}

{29.{~ }Jim nez, J.; Škalič, M.; Mart
nez-Rosell, G.; De Fabritiis, G., Kdeep: Protein CLigand Absolute
Binding Affinity Prediction Via 3d-Convolutional Neural Networks.
\emph{J. Chem. Inf. Model.} \textbf{2018}, 58, 287-296.}

{30.{ }Ragoza, M.; Hochuli, J.; Idrobo,
E.; Sunseri, J.; Koes, D. R., Protein CLigand Scoring with Convolutional
Neural Networks. \emph{J. Chem. Inf. Model.} \textbf{2017}, 57,
942-957.}

{31.{~ }Skalic, M.; Jim nez, J.; Sabbadin,
D.; De Fabritiis, G., Shape-Based Generative Modeling for De Novo Drug
Design. \emph{J. Chem. Inf. Model.} \textbf{2019}, 59, 1205-1214.}

{32.{~ }Golkov, V.; Skwark, M. J.;
Mirchev, A.; Dikov, G.; Geanes, A. R.; Mendenhall, J.; Meiler, J.;
Cremers, D., 3d Deep Learning for Biological Function Prediction from
Physical Fields. \emph{arXiv preprint arXiv:1704.04039} \textbf{2017}.}

{33.{~ }Winter, R.; Montanari, F.; No ,
F.; Clevert, D.-A., Learning Continuous and Data-Driven Molecular
Descriptors by Translating Equivalent Chemical Representations.
\emph{Chem. Sci.} \textbf{2019}, 10, 1692-1701.}

{34.{~ }Hinton, G. E.; Krizhevsky, A.;
Wang, S. D. Transforming Auto-Encoders. In Artificial Neural Networks
and Machine Learning C ICANN 2011, Berlin, Heidelberg, 2011; Honkela,
T.; Duch, W.; Girolami, M.; Kaski, S., Eds. Springer Berlin Heidelberg:
Berlin, Heidelberg, 2011; pp 44-51.}

{35.{~ }Sabour, S.; Frosst, N.; Hinton, G.
E. Dynamic Routing between Capsules. In Advances in Neural Information
Processing Systems, 2017; 2017; pp 3856-3866.}

{36.{~ }Sabour, S.; Frosst, N.; Hinton, G.
Matrix Capsules with Em Routing. In International Conference on Learning
Representations, 2018; 2018; pp 1-15.}

{37.{~ }Lenssen, J. E.; Fey, M.;
Libuschewski, P. Group Equivariant Capsule Networks. In Advances in
Neural Information Processing Systems, 2018; 2018; pp 8844-8853.}

{38.{~ }Venkatraman, S.; Balasubramanian,
S.; Sarma, R. R., Building Deep, Equivariant Capsule Networks.
\emph{arXiv preprint arXiv:1908.01300} \textbf{2019}.}

{39.{~ }Cramer, R. D.; Patterson, D. E.;
Bunce, J. D., Comparative Molecular Field Analysis (Comfa). 1. Effect of
Shape on Binding of Steroids to Carrier Proteins. \emph{J. Am. Chem.
Soc.} \textbf{1988}, 110, 5959-5967.}

{40.{~ }Klebe, G.; Abraham, U.; Mietzner,
T., Molecular Similarity Indices in a Comparative Analysis (Comsia) of
Drug Molecules to Correlate and Predict Their Biological Activity.
\emph{J. Med. Chem.} \textbf{1994}, 37, 4130-4146.}

{41.{~ }Halgren, T., Merck Molecular Force
Field. Iii. Molecular Geometries and Vibrational Frequencies for Mmff94.
\emph{J. Comput. Chem.} \textbf{1996}, 17, 553-586.}

{42.{~ }Halgren, T., Merck Molecular Force
Field. V. Extension of Mmff94 Using Experimental Data, Additional
Computational Data, and Empirical Rules. \emph{J. Comput. Chem.}
\textbf{1996}, 17, 616-641.}

{43.{~ }Halgren, T. A., Merck Molecular
Force Field. I. Basis, Form, Scope, Parameterization, and Performance of
Mmff94. \emph{J. Comput. Chem.} \textbf{1996}, 17, 490-519.}

{44.{~ }Halgren, T. A., Merck Molecular
Force Field. Ii. Mmff94 Van Der Waals and Electrostatic Parameters for
Intermolecular Interactions. \emph{J. Comput. Chem.} \textbf{1996}, 17,
520-552.}

{45.{~ }Tosco, P.; Stiefl, N.; Landrum,
G., Bringing the Mmff Force Field to the Rdkit: Implementation and
Validation. \emph{J. Cheminform.} \textbf{2014}, 6, 37.}

{46.{~ }Paszke, A.; Gross, S.; Massa, F.;
Lerer, A.; Bradbury, J.; Chanan, G.; Killeen, T.; Lin, Z.; Gimelshein,
N.; Antiga, L. Pytorch: An Imperative Style, High-Performance Deep
Learning Library. In Advances in Neural Information Processing Systems,
2019; 2019; pp 8024-8035.}

{47.{~ }Kingma, D. P.; Ba, J., Adam: A
Method for Stochastic Optimization. \emph{arXiv preprint
arXiv:1412.6980} \textbf{2014}.}

{48.{~ }Sterling, T.; Irwin, J. J., Zinc
15 C Ligand Discovery for Everyone. \emph{J. Chem. Inf. Model.}
\textbf{2015}, 55, 2324-2337.}

{49.{~ }Landrum, G. \emph{Rdkit:
Open-Source Cheminformatics}.}

{50.{~ }Lovrić, M.; Molero, J. M.; Kern,
R., Pyspark and Rdkit: Moving Towards Big Data in Cheminformatics.
\emph{Mol. Inform.} \textbf{2019}, 38, 1800082.}

{51.{~ }Davies, M.; Nowotka, M.;
Papadatos, G.; Dedman, N.; Gaulton, A.; Atkinson, F.; Bellis, L.;
Overington, J. P., Chembl Web Services: Streamlining Access to Drug
Discovery Data and Utilities. \emph{Nucleic Acids Res.} \textbf{2015},
43, W612-W620.}

{52.{~ }Gaulton, A.; Hersey, A.; Nowotka,
M.; Bento, A. P.; Chambers, J.; Mendez, D.; Mutowo, P.; Atkinson, F.;
Bellis, L. J.; Cibri n-Uhalte, E.; Davies, M.; Dedman, N.; Karlsson, A.;
Magariños, M. P.; Overington, J. P.; Papadatos, G.; Smit, I.; Leach, A.
R., The Chembl Database in 2017. \emph{Nucleic Acids Res.}
\textbf{2016}, 45, D945-D954.}

{53.{~ }\emph{Schrödinger Suites 2018-1},
Schrödinger, LLC, New York, NY, 2018.}

{54.{~ }Liu, Z.; Su, M.; Han, L.; Liu, J.;
Yang, Q.; Li, Y.; Wang, R., Forging the Basis for Developing Protein
CLigand Interaction Scoring Functions. \emph{Acc. Chem. Res.}
\textbf{2017}, 50, 302-309.}

{55.{~ }Hawkins, P. C. D.; Skillman, A.
G.; Warren, G. L.; Ellingson, B. A.; Stahl, M. T., Conformer Generation
with Omega: Algorithm and Validation Using High Quality Structures from
the Protein Databank and Cambridge Structural Database. \emph{J. Chem.
Inf. Model.} \textbf{2010}, 50, 572-584.}

{56.{~ }Gao, P.; Wang, L.; Zhao, L.;
Zhang, Q.-y.; Zeng, K.-w.; Zhao, M.-b.; Jiang, Y.; Tu, P.-f.; Guo,
X.-y., Anti-Inflammatory Quinoline Alkaloids from the Root Bark of
Dictamnus Dasycarpus. \emph{Phytochemistry} \textbf{2020}, 172, 112260.}

{57.{~ }Weill, N.; Rognan, D.,
Alignment-Free Ultra-High-Throughput Comparison of Druggable
Protein-Ligand Binding Sites. \emph{J. Chem. Inf. Model.} \textbf{2010},
50, 123-135.}

{58.{~ }Hu, G.; Kuang, G.; Xiao, W.; Li,
W.; Liu, G.; Tang, Y., Performance Evaluation of 2d Fingerprint and 3d
Shape Similarity Methods in Virtual Screening. \emph{J. Chem. Inf.
Model.} \textbf{2012}, 52, 1103-1113.}

{59.{~ }Awale, M.; Reymond, J.-L., Atom
Pair 2d-Fingerprints Perceive 3d-Molecular Shape and Pharmacophores for
Very Fast Virtual Screening of Zinc and Gdb-17. \emph{J. Chem. Inf.
Model.} \textbf{2014}, 54, 1892-1907.}

\end{document}